\documentclass[aoas,MSNbibl,nameyear,dvips]{arximspdf}
\usepackage{graphicx}
\doi{10.1214/12-AOAS562} %kopijuoti is PTS
\volume{6}
\issue{4}
\pubyear{2012}
\firstpage{1883}
\lastpage{1905}

\makeatletter
\newcommand{\thetavect}{\bolds{\theta}}
\newcommand{\pivect}{\bolds{\pi}}
\newcommand{\nuvect}{\bolds{\nu}}
\newcommand{\muvect}{\bolds{\mu}}

\newcommand{\fvect}{\mathbf{f}}
\newcommand{\yvect}{\mathbf{y}}

\newcommand{\T}{\mathrm{T}}
\newcommand{\Tr}{\operatorname{Tr}}
\newcommand{\E}{\mathrm{E}}

\newcommand{\Norm}{\mathcal{N}}
\makeatother

\begin{document}
\begin{frontmatter}

\title{Probabilistic prediction of neurological disorders with a
statistical assessment of neuroimaging data modalities}
\runtitle{Probabilistic prediction of neurological disorders}

\begin{aug}
\author[A]{\fnms{M.}~\snm{Filippone}\corref{}\ead[label=e1]{maurizio.filippone@glasgow.ac.uk}},
\author[B]{\fnms{A.~F.}~\snm{Marquand}\ead[label=e2]{andre.marquand@kcl.ac.uk}\thanksref{t1}},
\author[B]{\fnms{C.~R.~V.}~\snm{Blain}\ead[label=e3]{camillahertlein@googlemail.com}},
\author[B]{\fnms{S.~C.~R.}~\snm{Williams}\ead[label=e4]{steve.williams@kcl.ac.uk}},
\author[C]{\fnms{J.}~\snm{Mour\~{a}o-Miranda}\ead[label=e5]{j.mourao-miranda@cs.ucl.ac.uk}\thanksref{t2}}
\and
\author[D]{\fnms{M.}~\snm{Girolami}\thanksref{t3}\ead[label=e6]{mark@stats.ucl.ac.uk}\ead[label=e7]{third@somewhere.com}\ead[label=u1,url]{http://www.foo.com}}

\thankstext{t1}{Supported by the KCL Centre of Excellence in Medical
Engineering, funded by the Wellcome Trust and EPSRC under Grant WT088641/Z/09/Z.}
\thankstext{t2}{Supported by the Wellcome Trust under Grant WT086565/Z/08/Z.}
\runauthor{M. Filippone et al.}
\thankstext{t3}{Supported by the EPSRC Grants EP/E052029/1 and EP/H024875/1.}

\affiliation{University of Glasgow, King's College London, King's
College London, King's College London, King's College London and
University~College~London}
\address[A]{M. Filippone\\
School of Computing Science\\
University of Glasgow\\
Glasgow G12 8QQ\\
%Scotland\\
United Kingdom\\
\printead{e1}}% \phantom{E-mail:\ }\printead*{e2}

\address[B]{A.~F. Marquand\\
C.~R.~V. Blain\\
S.~C.~R. Williams\\
Institute of Psychiatry \\
King's College London\\
London\\
United Kingdom\\
\printead{e2}\\
\phantom{E-mail:\ }\printead*{e3}\\
\phantom{E-mail:\ }\printead*{e4}}

\address[C]{J. Mour\~{a}o-Miranda\\
University College and\\
King's College London\\
London\\
United Kingdom\\
\printead{e5}}

\address[D]{M. Girolami\\
Department of Statistical Science\\
Centre for Computational Statistics \\
\quad and Machine Learning \\
University College London\\
London\\
United Kingdom\\
\printead{e6}}

\end{aug}

% HISTORY:
\received{\smonth{10} \syear{2011}}
\revised{\smonth{4} \syear{2012}}

% ABSTRACT
%
\begin{abstract}
For many neurological disorders, prediction of disease state is an
important clinical aim.
% [afm] Neuroimaging measures provide a way to access information about
%the brain, but often presents particular challenges to statistical
%approaches in that the data is extremely high dimensional with very
%few samples.
Neuroimaging provides detailed information about brain structure and
function from which such predictions may be statistically derived.
%This is statistically challenging in that the data are extremely high
%dimensional and relatively few samples are available.
%Most techniques currently employed are nonprobabilistic
%classification methods based on a single neuroimaging modality and are
%unable, in any statistical manner, to fully address questions related
%to the importance of different sources of information derived from
%neuroimages.
A multinomial logit model with Gaussian process priors is proposed to:
(i) predict disease state based on whole-brain neuroimaging data and
(ii)~\mbox{analyze} the relative informativeness of different image modalities
and brain regions. Advanced Markov chain Monte Carlo methods are
employed to perform posterior inference over the model.
This paper reports a statistical assessment of multiple neuroimaging
modalities applied to the discrimination of three Parkinsonian
neurological disorders from one another and healthy controls, showing
promising predictive performance of disease states when compared to
nonprobabilistic classifiers based on multiple modalities.
The statistical analysis also quantifies the relative importance of
different neuroimaging measures and brain regions in discriminating
between these diseases
% [afm] and to give insights on the use of multiple sources for
%discriminating between classes.
% The findings reported in this study give important information on the
%relative importance of the different sources and different brain
%regions in line with what presented in the neuroimaging literature.
%The findings on the relative importance of the different imaging
%sequences for discrimination suggest
and suggests that for prediction there is little benefit in acquiring
multiple neuroimaging sequences.
Finally, the predictive capability of different brain regions is found
to be in accordance with the regional pathology of the diseases as
reported in the clinical literature.
\end{abstract}

% KEYWORDS
%

\end{frontmatter}
\newpage
%s1 #&#
\section{Introduction}\label{sec1}

For many neurological and psychiatric disorders, making a definitive
diagnosis and predicting clinical outcome are complex and difficult problems.
Difficulties arise due to many factors, including overlapping symptom
profiles, comorbidities in clinical populations and individual
variation in disease phenotype or disease course.
In addition, for many neurological disorders the diagnosis can only be
confirmed via analysis of brain tissue post-mortem.
Thus, technological advances that improve the efficiency or accuracy of
clinical assessments hold the potential to improve mainstream clinical
practice and to provide more cost-effective and personalized approaches
to treatment.

In this regard, combining data obtained from neuroimaging measures
% [AM] I will define this a bit later
with statistical discriminant analysis has recently attracted
substantial interest among the neuroimaging community [e.g., \citet
{Kloppel08}, \citet{Marquand08}].
% XMX
%
%
% For example, proof of concept applications have been demonstrated for
%automatically diagnosing schizophrenia \citep{Davatzikos05,Fan08c},
%depression \citep{Fu07,Marquand08,Hahn11}, Alzheimer's disease
% In the longer term, an even more important application of statistical
%methodology in clinical neuroimaging is for \textit{prognosis}, where
%they may be used to predict the outcome of a treatment before it is
%known or predict which individuals are likely to progress from a
%'prodromal' disease stage to a more advanced stage.
% For example, classification methods have been applied to brain images
%to predict whether depressed subjects will respond to treatment
%impairment will convert to Alzheimer's disease
%in Huntington's disease \citep{RizkJackson11}.
Neuroimaging data present particular statistical challenges in that
they are often extremely high dimensional (in the order of hundreds of
thousands to millions of variates) with very few samples (tens to hundreds).
{Further, multiple imaging sequences may be acquired for each
participant, each aiming to measure different properties of brain
tissue. Each sequence may in turn give rise to several different
measurements. In the present work, we will use the term ``modality'' to
describe such a set of measurements.}
In response to those challenges, most attempts to predict disease state
from neuroimaging data employ the Support Vector Machine [SVM; see,
e.g., \citet{Scholkopf01}] based on {information obtained from a single
modality}.
Such an approach, however, is not able, in any statistical manner, to
fully address questions related to the importance of different {modalities}.
{As we will show in the experimental section of this paper, even the
multi-modality SVM-based classifier proposed in \citet{Rakotomamonjy07}
lacks a systematic way of characterizing the uncertainty in the
predictions and in the assessment of the relative importance of
different modalities.}
% The SVM is regarded as having good predictive performance in high
%dimensional settings, but suffers from some limitations: first, it is
%fundamentally a binary classification device and extension to
%multiples classes of disease can be problematic. For many clinical
%problems, binary classification is an unrealistic oversimplification
%since there are potentially multiple disorders that share the same
%symptoms and any of the disorders may be present at the initial
%clinical consultation. Second, SVM does not lend itself easily to
%probabilistic prediction, which means that it does not quantify
%predictive uncertainty in a coherent manner. Accurate quantification
%of uncertainty is important so that the clinician can, for example,
%differentiate subjects having a high probability of having a disorder
%from more ambiguous cases. An additional difficulty with the approach
%outlined above is that it is often unknown in advance which
%neuroimaging measurements (e.g. different imaging sequences) or brain
%regions are optimal for discriminating between clinical conditions and
%whether combining multiple measurements will improve predictive
%performance by providing complementary information (e.g. by
%representing different aspects of disease pathology).
%%\blue{In the remainder of this paper, we will refer to data derived
%from different sources of neuroimaging data as modalities ; we will
%focus in particular on data obtained from different sources of
%neuroimaging data.}

% To address these limitations
In this work, we adopt a multinomial logit model based on Gaussian
process (GP) priors [\citet{Williams98}] as a probabilistic prediction
method that provides the means to incorporate measures from different
imaging modalities.
We apply this approach to discriminate between three relatively common
neurological disorders of the motor system based on data modalities
derived from three distinct neuroimaging sequences.
%The proposed classification model is an instantiation of latent
%Gaussian models (LGMs) \citep{Rue09}, where exact inference is
%intractable.
%Simple Laplace approximations \citep{Tierney86} could be used, but
%they would lead to very inaccurate predictions \citep{Kuss05}.
%
In this application we aim to characterize uncertainty without
resorting to potentially inaccurate deterministic approximations to the
integrals involved in the inference process.
Therefore, we propose to employ Markov chain Monte Carlo (MCMC) methods
to estimate the analytically intractable integrals, as they {provide
guarantees of asymptotic convergence} to the correct results.
The particular structure of the model and the large number of variables
involved, however, make the use of MCMC techniques seriously challenging
[\citet{FilipponeTECHREPLGMS12}, \citet{Murray10}].
In this work, we make use of reparametrization techniques [\citet{Yu11}]
and state-of-the-art sampling methods based on the geometry of the
underlying statistical model [\citet{Girolami11}] to achieve efficient sampling.

The remainder of the paper is structured as follows:
%% in section 2, we present a theoretical motivation for the approach
%employed,
%In section 4, we outline some variations to reduce computational
%complexity.
In Section~\ref{sec2} we describe the motivating application of statistical
discrimination of movement disorders from brain images. In Section~\ref{sec3} we
introduce the multinomial logit model with GP priors and in Section~\ref{sec4}
we present the associated MCMC methodology.
In Section~\ref{sec5} we report a comparison of MCMC strategies applied to our
brain imaging data and in Section~\ref{sec6} we investigate the predictive
ability of different data sources and brain regions, comparing the
results with a nonprobabilistic multi-modality classifier based on SVMs.
Section~\ref{sec7} shows how predictive probabilities can be used to refine
predictions, and Section~\ref{sec8} draws conclusions commenting on the
questions that this methodology addresses in this particular application.

%s2 #&#
\section{Discriminating among Parkinsonian disorders}\label{sec2}

%s2.1 #&#
\subsection{Introduction to the disorders}
For this application, we aim to discriminate between healthy control
subjects (HCs) and subjects with either multiple system atrophy (MSA),
progressive supranuclear palsy (PSP) or idiopathic Parkinson's disease
(IPD), which are behaviorally diagnosed motor conditions collectively
referred to as ``Parkinsonian disorders.''
MSA, PSP and IPD can be difficult to distinguish clinically in the
early stages [\citet{Litvan03}] and carry a high rate of misdiagnosis,
even though early diagnosis is important in predicting clinical outcome
and formulating a treatment strategy [\citet{Seppi07}].
For example, MSA and PSP have a much more rapid disease progression
relative to IPD and carry a shorter life expectancy after diagnosis.
Further, IPD responds relatively well to pharmacotherapy,
while MSA and PSP are both associated with a modest to poor response.
Thus, automated diagnostic tools to discriminate between the disorders
is of clinical relevance where they may help to reduce the rate of
misdiagnosis and ultimately lead to more favorable outcomes for patients.

%s2.2 #&#
\subsection{The clinical problem of discriminating among Parkinsonian
disorders}

In this study we employed magnetic resonance imaging (MRI), as it is
noninvasive, widely available and, unlike alternative measures such as
positron emission tomography, does not involve exposing subjects to
ionizing radiation. A detailed discussion of the imaging modalities
employed in this study is beyond the scope of the present work, but see
\citet{Farral06} for an overview. Briefly, the different imaging
modalities employed here measure different properties of brain tissue:
T1-weighted imaging is well-suited to visualizing anatomical structure,
while T2-weighted structural imaging often shows focal tissue
abnormalities more clearly. Diffusion tensor imaging (DTI) does not
measure brain structure directly, but instead measures the diffusion of
water molecules along fibre tracts in the brain, thus quantifying the
integrity of the fibre bundles that connect different brain regions
[see \citet{Basser02} for an introduction to DTI].

A review of the neuropathology of the Parkinsonian disorders is also
beyond the scope of this work, but briefly we note that MSA and PSP are
characterized by distinct cellular pathologies and subsequent
degeneration of widespread and partially overlapping brain regions.
For MSA, affected regions include the brainstem, basal ganglia (e.g.,
caudate and putamen), cerebellum and cerebral cortex [\citet
{Wenning97}]. Note that MSA is sometimes subdivided into Parkinsonian
and cerebellar subtypes (MSA-P and MSA-C, resp.), but for the
present work we included both variants in the same class. For PSP the
brainstem and basal ganglia both undergo severe degeneration [\citet
{Hauw94}], although cortical areas are also affected. In contrast, IPD
is characterized in the early stages by relatively focal pathology in
the substantia nigra (a pair of small nuclei in the brainstem), which
is difficult to detect using conventional structural MRI where the
scans of IPD patients can appear effectively normal [\citet{Seppi07}].

There is, however, some evidence that changes in IPD can be detected
using DTI [see, e.g., \citet{Yoshikawa04}]. Thus, it is of interest to
investigate which data modalities are best suited to discriminating
between MSA, PSP, IPD and HCs, which has practical implications in
planning future diagnostic studies: MRI scans are costly, so it is
desirable to know which data modalities provide the best discrimination
of the diagnostic groups and which scans can be omitted from a scanning
protocol to avoid wasting money acquiring scans that do not provide
additional predictive value.

%s2.3 #&#
\subsection{State of the art in diagnosis and prediction}

% \subsection{Comparison method}
We are aware of only one existing study that employed a discriminant
approach to diagnose these diseases based on whole-brain neuroimaging
measures [\citet{Focke11}]. This study employed binary SVM classifiers
to discriminate MSA-P from IPD, PSP and HCs based on a similar sample
to the present study. The authors reported that (i) PSP could be
accurately discriminated from IPD, (ii) that separation of the MSA-P
group from IPS and from controls was only marginally better than chance
and (iii)~that no separation of the IPD group from HCs was possible.
The authors did not attempt to combine the distinct binary classifiers
to provide multi-class predictions.
% [AM] I think it's best to leave this until the results section
%multi-class classifiers directly, the results presented in this work
%would appear to be an improvement with respect to those presented in

The problem of combining different modalities in classification models
can be seen as a Multiple Kernel Learning (MKL) problem [\citet
{Lanckriet04}, \citet{Sonnenburg06}].
A recent MKL approach to classification referred to as ``simpleMKL'' has
been proposed by \citet{Rakotomamonjy07}.
SimpleMKL is based on a SVM learning algorithm and shows good
performance relative to other MKL approaches; for this reason we will
consider it as a baseline against which we aim to compare the
performance of the proposed approach.

%s2.4 #&#
\subsection{Data acquisition and preprocessing}
Eighteen subjects with MSA, 16 subjects with PSP, 14 subjects with IPD
(all in mid-disease stage) and 14 HCs were recruited according to
clinical and experimental criteria described in \citet{Blain06}.
{For each subject, a T2-weighted structural image, a T1-weighted
spoiled gradient recalled (SPGR) structural image and a DTI sequence
were acquired and preprocessed (see the \hyperref[app]{Appendix} for the details on
acquisition).}

All images were screened by a trained radiologist and were examined for
gross structural abnormalities, including white matter abnormalities.
Diffusion tensor images were then preprocessed according to an in-house
protocol and were summarized by measures of fractional anisotropy (FA)
and mean diffusivity (MD) at every brain location [see \citet{Basser02}].
SPGR images were preprocessed using the DARTEL toolbox included in the
SPM software package (\href{http://www.fil.ion.ucl.ac.uk/spm}{www.fil.ion.ucl.ac.uk/spm}), which involved
nonlinear registration to a common reference space, segmentation into
grey matter (GM), white matter (WM) and cerebrospinal fluid (CSF) in
addition to smoothing with a 6mm isotropic Gaussian kernel.

\begin{figure}%[t]

\includegraphics{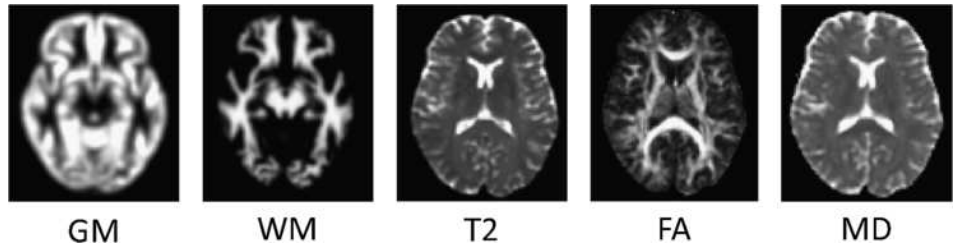}

\caption{Examples of each data source (after preprocessing), taken from
the same slice and subject.}
\label{figexamples}
\end{figure}

For this analysis, whole-brain (unmodulated) GM and WM images derived
from the SPGR scans, the T2 structural images plus the FA and MD images
derived from the DTI sequence were used for classification, yielding a
total of five distinct {modalities} for each subject.
For illustrative purposes, an example of each type of image after
preprocessing is provided in Figure~\ref{figexamples}.

%f1 #&#

%s3 #&#
\section{Multinomial logit model with GP priors}\label{sec3}

The aim of this study is twofold: the first is to reliably estimate
the probability that new subjects have MSA, PSP, IPD or none of them,
and the second is to assess the importance of different sources of
information in the discrimination among these diseases.
%As we will see shortly, in this work we will refer to modalities as
%data obtained from different neuroimaging sources or data from a
%single source but specific to brain regions.
We cast this problem as a multi-modality classification problem,
whereby we associate class labels corresponding to MSA, PSP, IPD and
HCs to $n$ subjects described by $s = 1, \ldots, q$ distinct
representations (i.e., modalities), each defined by $d_s$ covariates.

Denote each modality by an $n \times d_s$ matrix $X_s$.
Let $\{\yvect_1, \ldots, \yvect_n \}$ be the set of observed labels for
the $n$ subjects.
Assume a $1$-of-$m$ coding scheme ($m = 4$ in our application), whereby
the membership of subject $i$ to class $c$ is represented by a vector
$\yvect_i$ of length $m$ where $(\yvect_i)_r = 1$ if $r=c$ and zero otherwise.

% Consider data where a set $\{\yvect_1, \ldots, \yvect_n \}$ of $n$
%observed labels is associated with $n$ samples each described by a set
%of data sources. %imaging modalities. , that can be a set of regions
%or whole neuroimages.
% In this application samples are subjects and sources can be a set of
%whole-brain images from different imaging modalities or specific
%regions within them, and there are $s = 1, \ldots, q$ distinct
%representations (i.e. sources) for each sample, each defined by $d_s$
%covariates.
% Denote each source by an $n \times d_s$ matrix $X(s)$, and the number
%of distinct labels by $m$.
% We assume a $1$-of-$m$ coding scheme, whereby the membership of a
%sample $i$ to class $c$ is represented by a vector $\yvect_i$ of
%length $m$ where $(\yvect_i)_r = 1$ if $r=c$ and zero otherwise.
% In this application we are interested in modeling the probability $
%that a sample represented by $\mathbf{x}_i(1), \ldots,

In this work we propose a probabilistic multinomial logit
classification model based on Gaussian process (GP) priors to model the
probability $\pi_{ic} := p(y_{ic} = 1 | X_1, \ldots, X_q)$ that subject
$i$ belongs to class $c$.
The multinomial logit model assumes that the class labels $\yvect_i$
are conditionally independent given a set of $m$ latent functions
$\fvect_c$ that model the $\pi_{ic}$ using the following transformation:
%
%e1 #&#
\begin{equation}
\label{eqsoftmax} \pi_{ic} = \frac{[\exp(\fvect_c)]_i}{[\sum_{r=1}^{m} \exp(\fvect_r)]_i} .
\end{equation}

We assume that the latent variables $\fvect_c$ are independent across
classes and drawn from GPs, so that $\fvect_c \sim\Norm(\mathbf{0}, K_c)$.
The assumption of independence between variables belonging to different
classes can be relaxed in cases where there is prior knowledge about
that. %; this doesn't restrict the possibility to obtain posterior
%correlations between them after data are observed.
Note that the assumption of conditional independence of the class
labels given the latent variables is not restrictive, as the prior over
the latent variables imposes a covariance structure that is reflected
on the class labels.

In order to assess the importance of each modality in the
classification task, we propose to model each covariance $K_c$ as a
linear combination of covariances obtained from the $q$ modalities,
say, $C_s$ with $s = 1, \ldots, q$.
We constrain the linear combination of covariance functions to be
positive definite by modeling $K_c = \sum_{s=1}^{q} \exp[\theta_{cs}]
C_s$. %, so that inferring the weights will determine the relative
%importance of each source in the classification task.
Note that given the additivity of the linear model under the GP, it is
possible to interpret this model as one where each latent function is a
linear combination of basis functions with covariances $C_s$.
{Since the data modalities employed in this study potentially have
different numbers of features which are are also scaled differently, we
employed two simple operations to normalize the images prior to classification.
First we divided each feature vector by its Euclidean norm, then
standardized each feature to have zero mean and unit variance across
all scans.
We then chose a covariance for each modality to be $C_s = X_s X_s^T$.
% A linear correlation defined each covariance function such that $C_s
%= X_s X_s^T$.
% We inferred the weights of the linear combination of the covariance
%functions from each modality via the corresponding marginal posterior
%% $p({\boldsymbol\theta}|\textbf{y})$
% using the sampling methods described in section 3.1.
Given that the modalities are normalized and that the covariances are
linear in the data sources $X_s$, the inference of the corresponding
weights allows to draw conclusions on their relative importance in the
classification task.
In this work we imposed Gamma priors on the weights $\exp[\theta_{cs}]$,
but for the sake of completeness and to rule out any
dependencies of the results from the parametrization of the weights and
the specification of the prior, we have also explored the possibility
to use a Dirichlet prior inferring the concentration parameter; we will
discuss this in more detail in the sections reporting the results.

We note here that the representation in equation~(\ref{eqsoftmax}) is
redundant, as class probabilities are defined up to a scaling factor of
the exponential of the latent variables.
Choosing a model in which $m-1$ latent functions are modeled and one is
fixed would remove any redundancy, but, as described in \citet{Neal99},
``forcing the latent values for one of the classes to be zero would
introduce an arbitrary asymmetry into the prior.''
Also, modeling $m-1$ latent functions would not allow a direct
interpretation of the importance of different modalities given by the
hyper-parameters.

We used all features to perform the classification, because in our
experience feature selection does not provide a benefit in terms of
increasing the accuracy of classification models for neuroimaging data
but does increase their complexity.
In line with this, a recent comparative analysis of alternative data
preprocessing methods on a publicly available neuroimaging data set
indicated that feature selection did not improve classification
performance but did substantially increase the computation time [\citet
{Cuingnet10}].

% \blue{
% model the corresponding $n$-dimensional vector representation $
%a GP such that each $\boldsymbol{\phi}(s) \sim\Norm(\mathbf{0}, C_s)$,
%where the $n \times n$ covariance matrix has elements $i, j$ whose
%values represent the covariance between the $i$ and $j$ samples for
%source $s$.
% such that $f_{ci} = \sum_{s=1}^q \beta_{cs} \phi_{i}(s)$ where each
%function value, $\phi_{i}(s)$, is associated with the $s$-th source of
%the $i$-th sample.
% }

% Due to the additivity of the linear model under the GP and assuming
%independence between the $\boldsymbol{\phi}(s)$, then $\fvect_c \sim
%multinomial logit is then defined in the usual manner as
% \begin{equation} \label{eqsoftmax}
% \pi_{ic} = \frac{\exp(f_{ci})}{\sum_{r=1}^{m} \exp(f_{ri})}   .
% \end{equation}

% \blue{
% The dependence between class labels is given by the covariance
%structure of the GP prior on the latent variables, which is
%parametrized by a set of hyper-parameters.
% This model is therefore hierarchical with hyper-parameters
%conditioning the latent variables that condition the class labels.

We now discuss how to make inference for the proposed model.
In order to keep the notation uncluttered, we will drop the explicit
conditioning on~$X_s$.
We will denote by $\fvect$ the $(nc) \times1$ vector obtained by
concatenating the class specific latent functions $\mathbf{f}_c$, and
similarly by $\yvect$ and $\pivect$ the vectors obtained by
concatenating the vectors $\yvect_{\cdot c}$ and $\pi_{\cdot c}$.
Finally, let the $m \times q$ matrix $\thetavect$ denote the set of
hyper-parameters, and $K$ be the matrix obtained by block-concatenating
the covariance matrices $K_c$.

Given the likelihood for the observed labels, the prior over the latent
functions and the prior over the hyper-parameters, we can write the
log-joint density as
\[
\mathcal{L} = \log\bigl[p(\yvect, \fvect, \thetavect)\bigr] = \log\bigl[p(
\yvect| \fvect)\bigr] + \log\bigl[p(\fvect| \thetavect)\bigr] + \log\bigl[p(
\thetavect)\bigr] .
\]

One of the goals of our analysis is to obtain predictive distributions
for new subjects.
{Let $\yvect_*$ denote the corresponding label; the predictive density
is obtained by marginalizing out parameters and latent functions via}
\[
p(\yvect_* | \yvect) = \int p(\yvect_* | \fvect_*) p(\fvect_* | \fvect,
\thetavect) p(\fvect, \thetavect| \yvect) \,d\fvect_* \,d\fvect \,d\thetavect .
\]
In this work, we propose to estimate this integral by obtaining
posterior samples from $p(\fvect, \thetavect| \yvect)$ using MCMC methods.
The \hyperref[app]{Appendix} gives details of how Monte-Carlo estimates of this
predictive distribution may be obtained.
Obtaining samples from $p(\fvect, \thetavect| \yvect)$ is complex
because of the structure of the model that makes $\fvect$ and~$\thetavect$ strongly coupled a posteriori.
Also, there is no closed form for updating $\fvect$ and $\thetavect$
using a standard Gibbs sampler, so samplers based on an accept/reject
mechanism need to be employed (Metropolis-within-Gibbs samplers) with
the effect of reducing efficiency.
This motivates the use of efficient samplers to alleviate this problem
{as discussed next}.

%s4 #&#
\section{MCMC sampling strategies}\label{sec4}

%s4.1 #&#
\subsection{Riemann manifold MCMC methods}
The proposed model comprises a set of $m$ latent functions, each of
which has dimension $n$, and a set of $q \times m$ weights.
% Exact posterior inference for $p(\fvect, \thetavect| \yvect)$ is
%intractable under this model and we propose to carry out the inference
%task using MCMC methods.
Given the large number of strongly correlated variables involved in the
model, we need to employ statistically efficient sampling methods to
characterize the posterior distribution $p(\fvect, \thetavect| \yvect)$.

Recently, a set of novel Monte Carlo methods for efficient posterior
sampling has been proposed in \citet{Girolami11} which provides
promising capability for challenging and high-dimensional problems such
as the one considered in this paper.
%Consider MCMC methods based on accept/reject types of proposal, such
%as the Metropolis--Hastings (MH) algorithm
%of proposal mechanism, and Langevin Adjusted Metropolis Algorithm
%(MALA) \citep{Roberts98} and Hybrid Monte Carlo (HMC) \citep{Neal93}
%which are based on discrete Langevin diffusions and Hamiltonian
%mechanics respectively.
%%% offer an improvement given by the fact that gradients drive the
%exploration toward regions of higher density.
%In analogy with a physical system, in these methods it is necessary to
%specify the so called \textit{mass matrix}, whose size is quadratic in
%the number of model parameters.
In most sampling methods (with the exception of Gibbs sampling) it is
crucial to tune any parameters of the proposal distributions in order
to avoid strong dependency within the chain or the possibility that the
chain does not move at all.
As the dimensionality increases, this becomes a hugely challenging
issue, given that several parameters need to be tuned and are crucial
to the effectiveness of the sampler.

The sampling methods developed in \citet{Girolami11} aim at providing a
systematic way of designing such proposals by exploiting the
differential geometry of the underlying statistical model.
The main quantity in this differential geometric approach to MCMC is
the local Riemannian metric tensor which
% as defined in \citet{Girolami11}
is the expected Fisher Information (FI) that defines the statistical
manifold; see \citet{Girolami11} for full details.
The intuition behind manifold MCMC methods is that the statistical
manifold provides a structure that is suitable for making efficient
proposals based on Langevin diffusion or Hamiltonian dynamics.
% As an example, Hybrid Monte Carlo (HMC) allows to make proposals for
%the parameters $\thetavect$ by simulating a particle with mass matrix
%$M$ and momentum $\mathbf{p}$ moving in a potential field given by $-L(
% The variable $\mathbf{p}$ in an independent auxiliary variable with
%density $\Norm(\mathbf{p}|0, M)$ and gives rise to the factorization
%of the marginal densities for $\thetavect$ and $\mathbf{p}$.
% The negative log-joint density, that can be interpreted as a
%Hamiltonian for the particle, is
% $$
% H(\mathbf{p}) = -L(\thetavect) + \frac{1}{2} \log|M| + \frac{1}{2}
% $$
% Solving Hamilton's equations, we can simulate the motion of the
%particle; usually this is carried out by means of volume preserving
%integrators such as the leapfrog integration scheme.
% These integrators don't usually preserve the value of the Hamiltonian
%during the integration, and it is therefore necessary to accept the
%proposal using a Metropolis ratio in order to satisfy detailed balance.
In the case of the sampling of $\fvect$ using RM--HMC, let $G_\fvect$ be
the metric tensor computed as the FI for the statistical manifold of
$p(\yvect| \fvect)$ plus the negative Hessian of $p(\fvect|
\thetavect
)$ (see the \hyperref[app]{Appendix} for further details).
Introducing an auxiliary momentum variable $\mathbf{p} \sim\Norm
(\mathbf{p} | \mathbf{0}, G_{\fvect})$ as in HMC, RM--HMC can be derived
by solving the dynamics associated with the Hamiltonian:
\[
H(\fvect, \mathbf{p}) = -\log\bigl[p(\yvect, \fvect| \thetavect)\bigr] +
\tfrac
{1}{2} \log|G_{\fvect}| + \tfrac{1}{2} \mathbf{p}^{\T}
G_{\fvect}^{-1} \mathbf{p} + \mathrm{const.}
\]
Given that the metric tensor is dependent on the value of $\fvect$, the
Hamiltonian is therefore nonseparable between $\mathbf{p}$ and $\fvect$,
and a generalized leapfrog integrator must be employed [\citet{Girolami11}].
RM--HMC can be seen as a generalization of Hybrid Monte Carlo (HMC)
[\citet{Neal93}], where the mass matrix is now substituted by the
metric tensor.

% \citet{Girolami11} shows an application of this idea to MALA and HMC
%with the aim of devising efficient MCMC methods.

%Here we briefly review the basic notion of statistical manifold and FI.
%Consider a statistical model $S = \{ p(\yvect| \thetavect) |
%parameters $\thetavect$.
%Under some conditions that are satisfied for most commonly used models~
%called statistical manifold.
%Let $\mathcal{L} = \log[p(\yvect| \thetavect)]$; the FI matrix $G$ of
%$S$ at $\thetavect$ is defined as:
%$$
%G(\thetavect) = \E_{p(\yvect| \thetavect)} \left[\left(\nabla_{
%$$
%By definition, the FI matrix is positive semidefinite
%% $$
%% \sum_{i, j} c_i c_j g_{ij} = \E_{p(\yvect| \thetavect)} \left[
%% = \E_{p(\yvect| \thetavect)} \left[ \sum_{i} \left(c_i \frac{
%% $$
%and can be considered as the natural metric on $S$.
%
%Other key elements in information geometry are the Christoffel symbols
%that characterize connections on curved manifolds:
%$$
%$$

%s4.2 #&#
\subsection{Ancillary and sufficient augmentation}

The proposed classification model is hierarchical, and the application
of a Metropolis-within-Gibbs style scheme, sampling $\fvect|
\thetavect
, \yvect$ then $\thetavect| \fvect, \yvect$, leads to poor efficiency
and slow convergence.
This effect has drawn a lot of attention in the case of hierarchical
models in general [\citet{Papaspiliopoulos07}, \citet{Yu11}] and recently in
latent Gaussian models [\citet{Murray10}, \citet{FilipponeTECHREPLGMS12}]. %
In order to decouple the strong {posterior} dependency between
$\thetavect$ and $\fvect$, we can apply reparametrization techniques,
whereby we introduce a new set of variables $\nuvect_c$ related to the
old set of latent variables by a transformation $\fvect_c = g(\nuvect_c, \thetavect_c)$.
This transformation can be chosen to achieve faster convergence as
studied, for example, in \citet{Papaspiliopoulos07}, and should be
designed to handle both strong and weak data limits, namely, situations
where data overwhelm the prior or not.
In the terminology of \citet{Yu11}, we identify two particular cases,
namely, Sufficient augmentation (SA) and Ancillary augmentation (AA).
In the SA scheme the sampling of $\thetavect$ is done by proposing
$\thetavect^{\prime} | \thetavect, \fvect, \yvect$.
{In the case of weak data, as it is the case in this application}, the
SA parametrization is inefficient, given the strong coupling between
$\fvect$ and $\thetavect$.
{In contrast}, in the AA scheme, the new set of latent variables
$\nuvect_c$ is constructed to be a priori independent from $\thetavect
$; {this is a good candidate to provide an efficient parametrization in
cases of weak data}.
{To see this, in the case of no data, the posterior over
hyper-parameters and the newly defined latent variables $\nuvect_c$
corresponds to the prior which is factorized and easy to explore}.

This parametrization makes the $\nuvect_c$ ancillary for $\yvect$.
We propose to realize this by defining $\fvect_c = L_c \nuvect_c$,
where $L_c$ is any square root of $K_c$ (in the remainder of this paper
we will consider $L_c$ as the Cholesky factor of $K_c$).
This sampling scheme amounts to sampling $\thetavect^{\prime} |
\thetavect, \nuvect_1, \ldots, \nuvect_m, \yvect$.
%A combination of AA and SA to improve efficiency has been recently
%proposed in \citet{Yu11}.
In the next section we will report experiments showing the relative
merits of SA and AA combined with manifold methods, with the ultimate
goal of achieving efficiency in inferring latent functions and
hyper-parameters in our application.
All implemented methods were tested for correctness as proposed by
\citet{Geweke04}, and convergence analysis was performed using the
$\hat
{R}$ potential scale reduction factor [\citet{Gelman92}].

%t1 #&#
\begin{table}
\centering
\caption{Sampling schemes evaluated}\label{tabsamplers}
\begin{tabular*}{\textwidth}{@{\extracolsep{\fill}}lccccc@{}}
\hline
 & \multicolumn{2}{c}{$\bolds{p(\fvect'|\fvect,\thetavect)}$} &
\multicolumn{2}{c}{$\bolds{p(\thetavect'|\fvect,\thetavect)}$}& \\[-4pt]
 & \multicolumn{2}{c}{\hrulefill} &
\multicolumn{2}{c}{\hrulefill} &\\
\textbf{Approach}& \textbf{Sampler} & {\textbf{Metric}} & \textbf{Sampler} & {\textbf{Metric}} & {\textbf{Scheme}} \\
\hline
(a) & RM--HMC & $I$ & RM--HMC & $I$ & SA \\
(b) & RM--HMC & $\hat{F}$ & RM--HMC & $I$ & SA \\
(c) & RM--HMC & $G_{\fvect}$ & RM--HMC & $I$ & SA \\
(d) & RM--HMC & $G_{\fvect}$ & RM--HMC & $G_{\thetavect}$ & SA \\
(e) & RM--HMC & $\hat{F}$ & MH & -- & AA \\
(f) & RM--HMC & $G_{\fvect}$ & MH & -- & AA \\
\hline
\end{tabular*}
\end{table}

%s5 #&#
\section{Comparison of MCMC sampling strategies}\label{sec5}
In this section we investigate the efficiency of various MCMC sampling
strategies {in our application}.
Table~\ref{tabsamplers} lists the sampling approaches that we considered for this study.
All approaches make use of RM--HMC for sampling the latent variables
with different metrics.
Approach (a) uses a simple isotropic metric so that RM--HMC is
effectively HMC with an identity mass matrix.
Approaches (c), (d) and (f) use the metric derived from the FI (see
the \hyperref[app]{Appendix}), while approaches (b) and (e) use an alternative homogeneous
metric, defined as
$\hat{F} = K^{-1} + \operatorname{diag}(\pivect_p) - \Phi_p \Phi_p^{\T}$.
Note that\vadjust{\goodbreak} this is similar to the definition of the metric tensor
outlined in the \hyperref[app]{Appendix}, except $\pivect_p$ and $\Phi_p$ are defined
by the prior frequency of the classes in the training set instead of by
the likelihood. Employing this homogeneous metric is less efficient
than employing a position specific metric, but it holds two practical
advantages: (i) it has a substantially lower computational cost since
it does not recompute and invert the metric tensor at every step and
(ii) the explicit leapfrog integrator may be used in place of the
generalized (implicit) leapfrog integrator used in \citet{Girolami11}.

In sampling the hyper-parameters, approaches (a)--(c) effectively employ
an HMC proposal with identity mass matrix with SA parametrization.
Approach~(d) uses RM--HMC with metric derived from the FI as shown in
the \hyperref[app]{Appendix}, whereas approaches (e) and (f) make use of
Metropolis--Hastings (MH) with an identity covariance with AA parametrization.

Approach (a) can be viewed as a simple baseline approach since it does
not incorporate any knowledge {on the curvature} of the target
distribution and attempts to explore the parameter space by isotropic proposals.
It is presented primarily as a reference for the other approaches.
Approaches (b) and (c) employ manifold methods to efficiently sample
the latent variables only while approach (d) also applies them to
sample the hyper-parameters.
Approaches (e) and (f) employ manifold methods for the latent variables
and an MH sampler with AA for the hyper-parameters.

% \begin{table}
% \centering
% \caption{\label{tabsamplers} Sampling schemes evaluated}
% \begin{tabular}{{c}*{9}{c}}
% \hline
% Approach & \multicolumn{2}{c}{ $p(\fvect'|\fvect,\thetavect)$} &
% & Sampler & Variant & Sampler & Variant \\
% \hline
% (a) & RM--HMC & metric $ = I$ & RM--HMC & metric $ = I$ \\
% (b) & RM--HMC & metric $ = \hat{F}$ & RM--HMC & metric $ = I$ \\
% (c) & RM--HMC & metric $ = G_{\fvect}$ & RM--HMC & metric $ = I$ \\
% (d) & RM--HMC & metric $ = G_{\fvect}$ & RM--HMC & metric $ = G_{
% (e) & RM--HMC & metric $ = \hat{F}$ & MH & with AA \\
% (f) & RM--HMC & metric $ = G_{\fvect}$ & MH & with AA \\
% \hline
% \end{tabular}
% \end{table}

For all the experiments that follow, we applied an independent Gamma
prior to each weight $\exp(\theta_{cs})$, with $a = b = 2$, where $a$
and $b$ refer, respectively, to shape and rate parameters.
This prior is relatively vague but nevertheless constrained the sampler
to a plausible parameter range.
%% For the sake of completeness, we have also performed a series of
%tests by imposing a Dirichlet prior on the weights with concentration
%parameter $\alpha$, obtaining similar results, as discussed in section
%6.

We tuned each of the sampling approaches described above using pilot
runs and % XMX
% monitored convergence using Gelman and Rubin's potential scale
%reduction factor ($\hat{R}$; \citep{Gelman92}). We defined convergence
%as
assessed convergence by recording when all sampled variables had $\hat
{R} < 1.1$.
According to this criterion, sampling approaches (e) and (f) converged
after 1000 iterations for the latent function variables and after a
few thousands of iterations for the hyper-parameters. Sampling
approaches (a)--(d) did not converge even after 100,000 iterations, so
will not be considered further.
This demonstrates that the structure of the model poses a serious
challenge in efficiently sampling $\fvect$ and $\thetavect$, no matter
how efficient are the individual samplers employed in the
Metropolis-within-Gibbs sampler.
For all subsequent analysis, we discarded all samples acquired prior to
convergence (burn-in).
A plot reporting the evolution of Gelman and Rubin's shrink factor vs
the number of iterations (for the first 10,000 iterations) for the
slowest variable to converge is reported in Figure~\ref{figconvergenceplot}.
The left and right panel of Figure~\ref{figconvergenceplot} correspond
to the slowest variable in the approach (e) for the multi-modality and
multi-region classifiers (see next section), respectively; in both cases
the slowest variable was one of the hyper-parameters.

%f2 #&#
\begin{figure}

\includegraphics{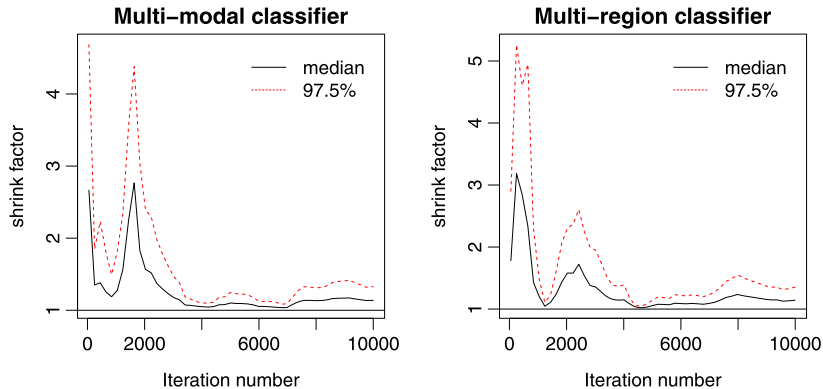}

\caption{Convergence analysis: plot reporting the evolution of Gelman
and Rubin's shrink factor vs the number of iterations for the slowest
variable to converge in approach \textup{(e)} for the multi-modality classifier
(left panel) and the multi-region classifier (right panel).}
\label{figconvergenceplot}
\end{figure}
%
%
%
%&
%
%

For the latent function variables, we used an RM--HMC trajectory length
of 10 leapfrog steps and a step size of 0.5 for sampling approaches (e)
and (f). This appeared to be near optimal and yielded an acceptance
rate in the range of 60--70\%, while keeping correlation between
successive samples relatively low. For the hyperparameters, a step size
of 0.2 yielded an acceptance rate in the range of 60--70\%, although
correlation between successive samples remained high (see below).

We report the Effective Sample Size (ESS) [\citet{Geyer92}] for each
method in Table~\ref{tabess1}, expressed as a percentage of the total
number of samples.
The ESS is an autocorrelation based method that is used to estimate
the number of independent samples within a set of samples obtained from
an MCMC method.
Both approaches (e) and (f) sampled the latent function variables
relatively efficiently, although there was some variability between
different variables.
% Upon closer inspection, the (RM-)HMC samplers were least efficient in
%the cases where one of the latent functions assumed the most extreme
%values. XMX
%
%t2 #&#
\begin{table}
\tablewidth=250pt
\caption{Efficiency of converged sampling schemes
(multi-source classifier). Min and max refer to the minimum and
maximum ESS across all sampled variables}\label{tabess1}
\begin{tabular*}{250pt}{@{\extracolsep{\fill}}lcc@{}}
% & & \em(min, max) & \em(min, max)\\
%
\hline
 & \textbf{Mean \%} $\bolds{\operatorname{ESS}_{f}}$& \textbf{Mean \%} $\bolds{\operatorname{ESS}_{\theta}}$ \\
\textbf{Approach}& \textbf{(min, max)} & \textbf{(min, max)} \\
\hline
(e) & 27.04 (5.31, 48.06) & 0.34 (0.21, 0.48) \\
(f) & 24.11 (5.71, 42.86) & 0.31 (0.19, 0.42) \\
\hline
\end{tabular*}
\end{table}
%
% \begin{table}
% \caption{\label{tabess1} Efficiency of converged sampling schemes
%(Multi-source classifier)}
% \fbox{
% \begin{tabular}{*{10}{c}}
% %\em Sampler ($f$) & \em Sampler ($\theta$) & \em mean \% $ESS_{f}$&
% % & & \em(min, max) & \em(min, max)\\
% %
% %\begin{tabular}{*{10}{c}}
% & \multicolumn{2}{c}{ $p(\fvect'|\fvect,\thetavect)$} &
%$ESS_{f}$& mean \% $ESS_{\theta}$ \\
% & \em Sampler & \em Variant & \em Sampler & \em Variant & (min, max)
%& (min, max) \\
% \hline
% (e) & RM--HMC & metric $= \hat{F}$ & MH & with AA & 27.04 (5.31,
%48.06) & 0.34 (0.21, 0.48) \\
% (f) & RM--HMC & metric $= G_{\thetavect}$ & MH & with AA & 24.11
%(5.71, 42.86) & 0.31 (0.19, 0.42) \\
% \end{tabular}}
% \end{table}
Sampling of the hyper-parameters was much more challenging than the
sampling of the latent function variables, and the MH samplers achieved
an ESS less than $0.5\%$ for all variables. Thus, for subsequent
analysis we ran a relatively long Markov chain (5~\mbox{million} iterations)
which we thinned by a factor of 500, ensuring independent sampling for
all variables.
Note that RM--HMC with metric $\hat{F}$ and RM--HMC with matrix
$G_{\fvect
}$ [approaches (e) and (f)] performed approximately equivalently for
sampling the latent function variables. Thus, for the remainder of this
paper we focus on the results obtained from the sampler that employed
the homogeneous metric $\hat{F}$ for the latent functions and MH for
the hyper-parameters [i.e., approach (e)], owing to its lower
computational cost.
% XMark
% We present trace plots (after thinning) for the first 1000 iterations
%of variables having: minimum, maximum and median ESS values in figure~
{The results reported in this section are in line with what is observed
in a recent extensive study on the fully Bayesian treatment of models
involving GP priors [\citet{FilipponeTECHREPLGMS12}].
In particular, it has been reported that the sampling of the latent
variables can be done efficiently using RM--HMC and a variant with the
homogeneous metric $\hat{F}$, and that for the hyper-parameters the MH
proposal with the AA parametrization is a good compromise between
efficiency and computational cost.}

%s6 #&#
\section{Predictive accuracy and assessment of neuroimaging data modalities}\label{sec6}

In this section we have three main objectives: first, we aim to
demonstrate that the predictive approach we propose can accurately
discriminate between multiple neurological conditions. Second, we
investigate which neuroimaging data modalities carry discriminating
information for these disorders and whether greater predictive
performance can be achieved by combining multiple modalities. Finally,
we investigate the predictive ability of different brain regions for
discriminating between each of the disorders.

For estimating the predictive ability of the classifiers we performed
four-fold cross-validation (CV).
In the CV procedure, we randomly partitioned the data into four folds
so that each CV fold contained approximately the same frequency of
classes as in the entire data set.
We then carried out the inference leaving out one fold that we used to
assess the accuracy of the proposed method; leaving out one fold at a
time, it is possible to obtain an estimate of performance on unseen data.

We compared the performance of the proposed multinomial logit model
with simpleMKL.
Similar to the proposed method, simpleMKL allows an optimal linear
combination of data sources or brain regions to be {inferred} from the
data, but, unlike the proposed approach, simpleMKL is not a
probabilistic model. In the MKL literature, each data source is
referred to as a ``kernel'' which corresponds to a covariance function
for the proposed multinomial logit model.
Since SVMs do not support true multi-class classification, we employed
a ``one-vs-all'' approach to combine multiple binary classifiers to
provide a multi-class decision function. This has the consequence that
simpleMKL estimates a linear combination of kernels that provides
optimal accuracy across all binary classification decisions, and is
therefore not able to estimate an independent set of kernel weighting
factors for each class. To ensure the comparison with the multinomial
logit model was as fair as possible, we used nested cross-validation to
find an optimal value for the SVM regularization parameter C. We
achieved this by performing an inner ``leave-one-out'' cross validation
cycle (``validation'') within each outer four-fold cycle (``test'') while
we varied C logarithmically across a wide range of values ($10^{-5}$ to
$10^5$ in steps of $10$). We selected the value of C that provided the
optimal accuracy on the validation set, before applying it to the test
set. To further examine whether any performance difference could be
attributed to the extension of simpleMKL to multiclass classification,
we also compared the accuracy of simpleMKL and the proposed model on
all possible binary classification decisions.
For simpleMKL, we used the implementation provided by \citet
{Rakotomamonjy07} where we used the ``weight variation'' stopping
criterion and the default options.

We employed two measures of predictive performance: (i) balanced
classification accuracy, which measures the mean number of correct
predictions across all classes assuming a zero--one loss and (ii) a
multi-class Brier score, which also quantifies the confidence of
classifier predictions on $w$ unseen samples and can be computed as the
following error measure: $ B = \frac{1}{w}\sum_{i=1}^{w} \sum_{c=1}^{m}
(\pi^*_{ic} - y_{ic}^*)^2$.
Note that simpleMKL does not provide probabilistic predictions, so the
Brier score is not appropriate to evaluate the performance of this algorithm.
%Note that the Brier score is an error measure, meaning that lower
%values imply better predictive performance. XMX
{\citet{Gneiting07} reported studies on the connections between the
Brier score and predictive accuracy in the case of two-class
classification, reporting that simple accuracy is a proper score unlike
the Brier score which is strictly proper.
We are unaware of any results on the connections between the two scores
in the case of multi-class classification.}

For comparison we also present predictive accuracy measures derived
from classifiers using each data source independently, and a classifier
using an unweighted linear sum of data sources (i.e., $K_c = \sum_{s=1}^{q} C_s$ for all $c$).

%s6.1 #&#
\subsection{Multi-modal classifier}

We first studied the classification problem based on the five data
sources obtained from the three modalities, namely, GM, WM, T2, FA and
MD, as explained in Section~\ref{sec2}, so that $q=5$.
The overall performance of each model is summarized in Table~\ref
{tabacc}. %% and a confusion matrix for the weighted sum classifier is
%provided in supplementary material.
Note that all classifiers exceeded the predictive accuracy that would
be expected by chance (i.e., $p < 0.05$, $\chi^2$ test).

% I will keep three significant digits only XMX
%
%t3 #&#
\begin{table}
\caption{Predictive accuracy (multi-source classifier).
Min and max values refer to minimum and maximum values across CV folds}\label{tabacc}
\begin{tabular*}{\textwidth}{@{\extracolsep{\fill}}lccc@{}}
\hline
& \textbf{Input data} & \textbf{Accuracy (min, max)} & \textbf{Brier score (min, max)} \\
\hline
1 & GM only & 0.627 (0.321, 0.854) & 0.667 (0.636, 0.712) \\
2 & WM only & 0.603 (0.350, 0.771) & 0.653 (0.609, 0.710) \\
3 & T2 only & 0.545 (0.500, 0.604) & 0.663 (0.619, 0.695) \\
4 & FA only & 0.569 (0.442, 0.688) & 0.675 (0.645, 0.703) \\
5 & MD only & 0.623 (0.533, 0.750) & 0.631 (0.588, 0.680) \\
6 & Weighted sum & 0.598 (0.350, 0.708) & 0.588 (0.517, 0.662) \\
7 & Unweighted sum & 0.610 (0.400, 0.708) & 0.553 (0.469, 0.646) \\
8 & SimpleMKL & 0.418 (0.143, 0.625) & -- \\
\hline
\end{tabular*}
\end{table}

% \begin{table}
% \caption{\label{tabacc} Predictive accuracy (multi-source classifier)}
% \centering
% \fbox{
% \begin{tabular}{*{10}{c}}
% & \em Input data & \em Accuracy (min, max) & Brier score (min, max) \\
% \hline
% 1 & GM only & 0.6271 (0.3208, 0.8542) & 0.6671 (0.6355, 0.7120) \\
% 2 & WM only & 0.6031 (0.3500, 0.7708) & 0.6531 (0.6092, 0.7102) \\
% 3 & T2 only & 0.5448 (0.5000, 0.6042) & 0.5688 (0.6193, 0.6945) \\
% 4 & FA only & 0.5688 (0.4417, 0.6875) & 0.6751 (0.6452, 0.7034) \\
% 5 & MD only & 0.6229 (0.5333, 0.7500) & 0.6314 (0.5877, 0.6798) \\
% 6 & Weighted sum & 0.5979 (0.3500, 0.7083) & 0.5857 (0.5169, 0.6615)
% 7 & Unweighted sum & 0.6104 (0.4000, 0.7083) & 0.5534 (0.4688,
%0.6462) \\
% \end{tabular}}
% \end{table}

%
%%\includegraphics[width=1
%

From Table~\ref{tabacc}, it is apparent that classifiers based on the
T2 and FA data sources achieved lower classification accuracy than all
the other data sources, suggesting they are not ideally suited to
discriminating between these disorders. Further, the linear
combinations of sources did not achieve higher accuracy than the best
individual data source and the highest accuracy was obtained using the
GM images only, although the difference is relatively small. The
simpleMKL classifier produced lower accuracy than either linear
combination derived from the multinomial logit model. {The mean
accuracy for the binary classifiers over all pairs of classes was
slightly higher for the GP classifiers (0.807) relative to simpleMKL
(0.741), suggesting that most of the performance difference between
simpleMKL and the proposed multinomial logit approach can be attributed
to the extension of simpleMKL to the multi-class case.}

In contrast to the outcomes for classification accuracy, the linear
combinations of data sources produced more accurate probabilistic
predictions than any of the individual modalities, indicative of a
disjunction between categorical classification accuracy and accurate
quantification of predictive uncertainty (Table~\ref{tabacc}). This is probably a
result of this model having greater flexibility to scale the magnitude
of the latent function variables.

%f3 #&#
\begin{figure}
\centering
\begin{tabular}{@{}cc@{}}

\includegraphics{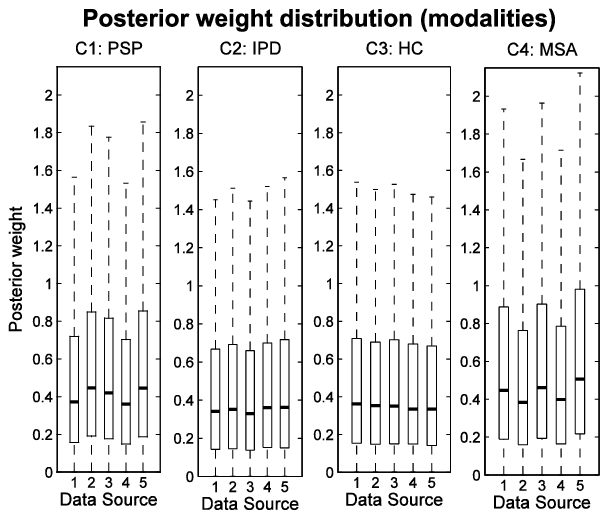}
 & \includegraphics{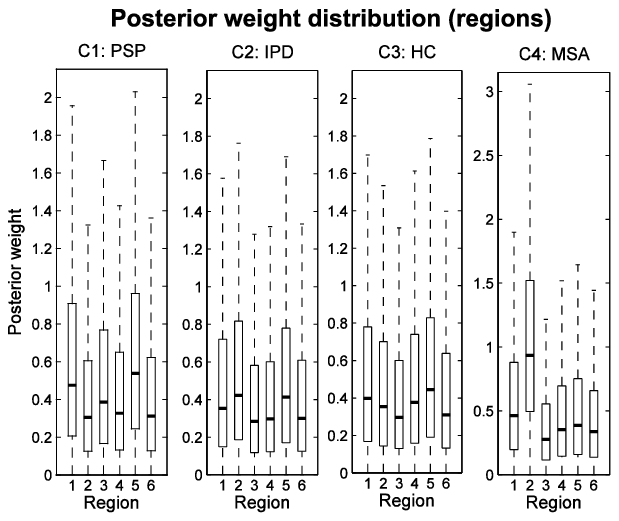}\\
\footnotesize{(a)} & \footnotesize{(b)}\\
\end{tabular}
\caption{Posterior distributions for the predictive weights for the
multi-modality classifier
\textup{(a)} and the multi-region classifier \textup{(b)}. Panel \textup{(a)}:
Data sources: (1) GM, (2) WM, (3) T2, (4) FA, (5) MD. Panel~\textup{(b)}:
Regions: (1) brainstem, (2) cerebellum, (3) caudate, (4) middle
occipital gyrus, (5) putamen, (6)~all other regions.}
\label{accbyclass1}
\end{figure}

%s6.1.1 #&#
\subsubsection{Covariance parameters for the latent functions}

The posterior distribution of the weights is an important secondary
outcome from this model and is summarized in Figure~\ref{accbyclass1}(a).
These hyper-parameters collectively describe the relative contribution
(or weighting factors) for each modality in deriving the prediction for
each class.

The posterior class distribution for covariance weights in this model
is relatively flat across all modalities for each class although each
weight has slightly greater magnitude for the PSP and MSA classes
relative to the other classes. Overall, the results from this section
provide evidence that all imaging modalities contain similar
information for discriminating disease groups. In other words, we found
little benefit from combining multiple neuroimaging sequences. This has
the important implication that for the purposes of discrimination it
appears sufficient to acquire a single structural MRI scan (i.e., SPGR
image), which is comparatively rapid and inexpensive to acquire.
Although the T2 images are also relatively inexpensive, they do not
offer the same discriminative value and the DTI images, which are
time-consuming and expensive to acquire, and appear to offer little
additional benefit.

%f4 #&#
\begin{figure}

\includegraphics{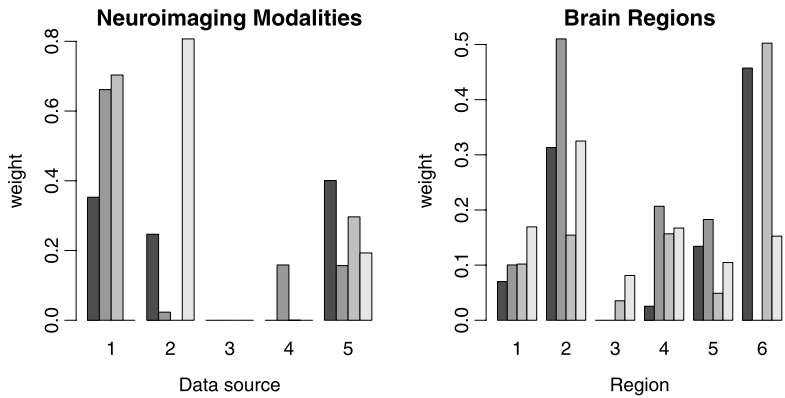}

\caption{Weights of the neuroimaging modalities (left panel) and brain
regions (right panel) obtained by simpleMKL across the four folds.
{Left panel: Data sources: (1) GM, (2) WM, (3) T2, (4)~FA, (5) MD.
Right panel: Regions: (1) brainstem, (2) cerebellum, (3) caudate, (4)~middle occipital gyrus, (5) putamen, (6) all other regions.}}
\label{weightssimplemkl}
\end{figure}

In the left panel of Figure~\ref{weightssimplemkl} we report the kernel
weights obtained by simpleMKL across the four folds.
We can see how the values of the weights are not consistent across the
folds; for example, GM is given zero weight in the fourth fold, whereas
it seems to be important for the other three cases.
Modality T2, instead, is consistently given zero weight across the four
folds, suggesting that this might not add any information to the other
modalities.
This is in contrast with the results of the probabilistic classifier
that suggests that there is not much evidence in the data to completely
ignore the information from one of the modalities.

%%%%%%%%%%%%%%%%%%%%%%%%%%%%%%%%%%%%%%%%%%%%%%%%%%%%%%%%%%%%%%%%%%%%%%
%%%%%%%%%%%%%%%%%%% Single modality classification %%%%%%%%%%%%%%%%%%%
%%%%%%%%%%%%%%%%%%%%%%%%%%%%%%%%%%%%%%%%%%%%%%%%%%%%%%%%%%%%%%%%%%%%%%

%s6.2 #&#
\subsection{Multi-region classifier}

In this section we illustrate how the proposed methodology may be used
to estimate the predictive value of different brain regions for
classification, although we will investigate the relative contribution
of different brain regions in greater detail and comment on the
clinical significance of these findings in a separate report. For
neurological applications, it is primarily important to assist
interpretation, since it is desirable to identify differential patterns
of regional pathology for each disease. While there are other methods
to achieve this goal, an advantage of the proposed approach is that it
provides a full posterior distribution over regional weighting
parameters. For this analysis we used only the GM data modality, since
it showed the highest discrimination accuracy, and used an anatomical
template [\citet{Shattuck08}] to parcellate the GM images into six
regions: (i) brainstem, (ii) bilateral cerebellum, (iii) bilateral
caudate, (iv) bilateral middle occipital gyrus, (v) bilateral putamen
and (vi) all other regions, so that now $q=6$. As described above, the
cerebellum, brainstem, caudate and putamen are affected to varying
degrees in MSA, PSP or IPD. The middle occipital gyrus region was
selected as a control region, as this is hypothesized to contain
minimal discriminatory information.
% Thus, for this analysis we associated sources to brain regions and we
%repeated the same experimental analysis. % XMX
% employ the same feature standardisation methods described above.
%These covariance matrices were then used as input to the classifier
%and a linear combination of regions was estimated in the same manner
%as described above. A similar four-fold cross-validation approach was
%employed to ensure generalizability.

% \subsection{Inferring the posterior distributions (multi-region
%classifier)} XMX

% For this analysis we re-evaluated all sampling methods noted above,
%which were also tuned similarly so that acceptance ratios were in the
%range of 50--70\% for both latent functions and hyperparameters. XMX
All sampling approaches performed similarly for this problem in that
none of the sampling approaches (a--d) converged after 100,000
iterations and sampling approaches (e) and (f) converged after 1000
iterations for the latent function variables and after a few thousands
of iterations for the hyper-parameters (see the right panel of Figure
\ref{figconvergenceplot}).
{Table~\ref{tabess2} reports the efficiency of the sampling approaches
(e) and (f), as they were the only ones that converged in a reasonable
number of iterations.}

%t4 #&#
\begin{table}
\tablewidth=250pt
\caption{Efficiency of converged sampling schemes
(multi-region classifier). Min and max refer to the minimum and
maximum ESS across all sampled variables}\label{tabess2}
\begin{tabular*}{250pt}{@{\extracolsep{\fill}}lcc@{}}
\hline
& \textbf{Mean \%} $\bolds{\operatorname{ESS}_{f}}$& \textbf{Mean \%} $\bolds{\operatorname{ESS}_{\theta}}$ \\
\textbf{Approach} & \textbf{(min, max)} & \textbf{(min, max)} \\
\hline
(e) & 8.52 (2.37, 34.85) & 0.28 (0.11, 0.64) \\
(f) & 9.42 (2.32, 35.44) & 0.31 (0.11, 0.50) \\
\hline
\end{tabular*}
\end{table}

% I will keep three significant digits only XMX
%
%t5 #&#
\begin{table}[b]
\caption{Predictive accuracy (multi-region classifier).
Min and max values refer to minimum and maximum values across CV folds}\label{tabacc2}
\begin{tabular*}{\textwidth}{@{\extracolsep{\fill}}lccc@{}}
\hline
& \textbf{Input data} & \textbf{Accuracy (min, max)} & \textbf{Brier score (min, max)} \\
\hline
1 & Brainstem only & 0.578 (0.500, 0.688) & 0.595 (0.555, 0.643) \\
2 & Cerebellum only & 0.478 (0.333, 0.562) & 0.634 (0.593, 0.643) \\
3 & Caudate only & 0.349 (0.221, 0.520) & 0.737 (0.693, 0.764) \\
4 & Mid. occipital gyrus only & 0.419 (0.333, 0.479) & 0.741 (0.677,
0.773) \\
5 & Putamen only & 0.438 (0.354, 0.521) & 0.668 (0.604, 0.718) \\
6 & All other regions & 0.424 (0.321, 0.563) & 0.753 (0.724, 0.779) \\
7 & Weighted sum & 0.614 (0.500, 0.708) & 0.547 (0.499, 0.593) \\
8 & Unweighted sum & 0.624 (0.500, 0.708) & 0.546 (0.501, 0.592) \\
9 & simpleMKL & 0.229 (0.111, 0.375) & -- \\
\hline
\end{tabular*}
\end{table}

% \begin{table}
% \caption{\label{tabess2} Efficiency of converged sampling schemes
%(multi-region classifier)}
% \fbox{
% \begin{tabular}{*{10}{c}}
% & \multicolumn{2}{c}{ $p(\fvect'|\fvect,\thetavect)$} &
%$ESS_{f}$& mean \% $ESS_{\theta}$ \\
% & \em Sampler & \em Variant & \em Sampler & \em Variant & (min, max)
%& (min, max) \\
% \hline
% (e) & RM--HMC & metric $= \hat{F}$ & MH & with AA & 8.52 (2.37, 34.85)
%& 0.28 (0.11, 0.64) \\
% (f) & RM--HMC & metric $= G_{\thetavect}$ & MH & with AA & 9.42 (2.32,
%35.44) & 0.31 (0.11, 0.50) \\
% \end{tabular}}
% \end{table}

The sampling efficiency for the latent function values was somewhat
lower for this problem than for the multi-modal prediction problem
described in the previous section.
% On closer inspection, this was due to the hyperparameters having
%greater variability in their sampled values.
On average, the sampling efficiency for the hyper-parameters was
approximately equivalent to the values reported above, but the minimum
ESS was slightly lower.\vadjust{\goodbreak}
To accommodate this, we thinned all Markov chains by a factor of 1000,
ensuring approximately independent sampling for all variables.

Predictive accuracy measures for the multi-region classifier are
presented in Table~\ref{tabacc2}. % and a confusion matrix for the
%weighted sum classifier is provided in supplementary material.
% Once again, all classifiers exceeded the predictive accuracy that
%would be expected by chance XMX
All classifiers exceeded the predictive accuracy that would be expected
by chance ($p < 0.05$, $\chi^2$ test) except the simpleMKL classifier
which performed very poorly for this data set.
{As for the multi-modal classifier, we compared the predictive accuracy
of simpleMKL to the logit model across all binary classification
decisions. In this case, the models produced similar accuracy (0.765
for the GP classifiers, 0.780 for simpleMKL). This provides further
evidence that the suboptimal performances of simpleMKL can be traced to
the extension of the binary SVM to the multi-class setting. In
particular, it is likely that the suboptimal performance of simpleMKL
in the multi-class context is due to the fact that} it does not support
different weighting factors for each class. In this case, the
classifiers using weighted and unweighted covariance sums of brain
regions produced the most accurate predictions and quantified
predictive confidence most accurately. Again, there was negligible
difference between the classifiers using the weighted and unweighted sums.

% \begin{table}
% \caption{\label{tabacc2} Predictive accuracy (multi-region
%classifier)}
% \centering
% \fbox{
% \begin{tabular}{*{10}{c}}
% & \em Input data & \em Accuracy (min, max) & Brier score (min, max) \\
% \hline
% 1 & Brainstem only & 0.5781 (0.5000, 0.6875) & 0.5950 (0.5545,
%0.6428) \\
% 2 & Cerebellum only & 0.4782 (0.3333, 0.5625) & 0.6337 (0.5929,
%0.6431) \\
% 3 & Caudate only & 0.3490 (0.2209, 0.5208) & 0.7365 (0.6929, 0.7642)
% 4 & Mid. Occipital Gyrus only & 0.4188 (0.3333, 0.4792) & 0.7413
%(0.6771, 0.7730) \\
% 5 & Putamen only & 0.4375 (0.3542, 0.5208) & 0.6681 (0.6041, 0.7178)
% 6 & All other regions & 0.4240 (0.3208, 0.5625) & 0.7528 (0.7244,
%0.7788) \\
% 7 & Weighted sum & 0.6135 (0.5000, 0.7083) & 0.5467 (0.4994, 0.5932)
% 8 & Unweighted sum & 0.6240 (0.5000, 0.7083) & 0.5460 (0.5007,
%0.5917) \\
% \end{tabular}}
% \end{table}

%%{acc_by_class2.eps}
%brainstem, (2) cerebellum, (3) caudate, (4) middle occipital gyrus,
%(5) putamen, (6) all other regions}}

%s6.2.1 #&#
\subsubsection{Covariance parameters for the latent functions}

The posterior distribution of the weights for the multi-region
classifiers is summarized in Figure~\ref{accbyclass1}(b).
%Again, this shows a correspondence with the accuracies reported above
%in that regions that discriminated a given class more accurately
%tended to be weighted more strongly.
The posterior means of the weighting factors were again relatively
constant between brain regions and also showed a high variance. This
indicates that the relative contribution of different brain regions was
not strongly determined by the data and that we should be cautious
about interpreting the relative contributions of the different brain
regions using this approach.
% XMark
% Further, it suggests that alternative methods such as Bayesian
%multinomial logistic regression may be more appropriate to determine
%the relative contribution of different brain regions.
Nevertheless, the clearest differential effect among regions was for
the cerebellum, where the lower quartile of the posterior distribution
for the MSA class was greater than the mean of all other regions. In
addition, the brainstem also made a small positive contribution toward
predicting the MSA class. As described above, both the cerebellum and
brainstem are known to undergo severe degeneration in MSA. The
strongest positive contributions to predicting the PSP class were
obtained from the brainstem, caudate and putamen, which once again are
regions known to show the extensive degeneration in PSP. The regional
weighting factors for the IPD and control groups were somewhat flatter,
which is consistent with the focal nature of the degeneration in
early-mid IPD and with the observation that the brain scans of these
groups are difficult to discriminate from one another. However, the
posterior suggests that the cerebellum showed a relatively increased
weighting relative to other regions for the IPD class, and that the
putamen was assigned a relatively increased weighting for the IPD and
HC classes, which is congruent with the expectation that these classes
are characterized by greater GM concentration in those regions relative
to the PSP and MSA classes respectively. From the current analysis, it
is difficult to determine the regions having the greatest predictive
value for discriminating the PD from the HC group. As future work,
separate binary classifiers trained to discriminate these classes
directly may be beneficial in this respect. We notice also that the
control region (i.e., the middle occipital gyrus) was assigned
comparatively low weighting for every class.

%scale for class 4}

Overall, the results from this section suggest that distributed
patterns of abnormality across multiple brain regions are necessary to
accurately discriminate between classes. The neurodegenerative
disorders studied in the present work have relatively well-defined
regional pathology, but even in this case the most accurate predictions
were obtained from the classifiers using all brain regions.

Again, the analysis of the weights obtained by simpleMKL (right panel
of Figure~\ref{weightssimplemkl}) shows that the nonprobabilistic
classifier obtains sparse solutions for the weights that are not
consistent across the folds, thus preventing one from being able to
properly assess the role played by each region in the classification task.

%s6.3 #&#
\subsection{Results with the Dirichlet prior}
Here we briefly discuss the results obtained when imposing a Dirichlet\vadjust{\goodbreak}
prior on the weights $\exp(\theta_{cs})$, focusing only on the
multi-region classifier for the sake of brevity.
The sampling strategy was as in approach (e), with the difference that
the update of the hyper-parameters followed a MH sampling with proposal
based on Dirichlet distributions.
In order to test the robustness to prior specification, we added a
further level in the hierarchy of the model by imposing a prior over
the concentration parameter of the Dirichlet distribution, so that the
model had a joint density $p(\yvect, \fvect, \thetavect, \alpha) =
p(\yvect| \fvect) p(\fvect| \thetavect) p(\thetavect| \alpha)
p(\alpha)$.
Including a hyper-prior over the concentration parameter has the effect
of averaging out the choice of the prior and inferring the
concentration parameter allows inference of levels of sparsity from the data.

% Most importantly, the posterior over the concentration parameter
%allows to infer the sparsity of the combination from data.
% The effect of reduction in the information gained at each level of
%the hierarchy has been studied in detail in , and although it is not
%true in general for all information measures, it can be noticed in
%most cases.

From the computational perspective, the sampling of $\alpha$ induces a
further level of correlation in the chains.
We ran thorough convergence tests, and we obtained similar convergence
and efficiency results as in the previous analysis.
We chose a fairly diffuse hyper-prior $p(\alpha)$ as exponential with
unit rate, so that $\E[\alpha] = 1$, which corresponds to a uniform
prior over the simplex for the weights.

Note that data has quite a weak effect in informing the posterior over
the concentration parameter, as they are three levels apart in the
hierarchy [\citet{Goel81}].
Nevertheless, comparing prior and posterior over $\alpha$, we notice a
slight reduction in the interquartile range from $[0.29, 1.39]$ to
$[0.51, 1.49]$ and a shift of the mean from $1$ to $1.16$, thus
supporting a diffuse (nonsparse) prior over the weights. %% (see Fig.
In terms of questions addressed in this particular application, the
results obtained by adding a hyper-prior lead to the same conclusions.
%% (see figure 1 in the supplementary material).
% The convergence analysis and the conclusions from the inference
%turned out to be very similar to those obtained in the case of Gamma
%priors on the weights; such results can be seen in the supplementary
%material.

%s7 #&#
\section{Refining predictions using predictive probabilities}\label{sec7}
We have discussed how a probabilistic approach allows us to assess the
importance of different neuroimaging modalities in disease
classification. Another advantage of employing a probabilistic
classification model is that predictive probabilities quantify the
uncertainty in the outcome, which allow a ``reject option'' to be
specified. Under this framework, the researcher specifies in advance a
confidence threshold below which a prediction is considered to be
inconclusive. In cases where the maximum class probability does not
exceed this threshold, the final decision may be deferred to another
classification model or a human expert. To investigate the suitability
of the proposed classifier for this approach and to assess the accuracy
of the classifier across varying rejection thresholds, we plotted
accuracy--reject curves for each of the classifiers investigated in this
work (Figure~\ref{accreject}). These were constructed by varying the
rejection threshold monotonically from 0 to 1 in steps of 0.01. At each
threshold, we computed the rejection rate as the proportion of samples
for which the most confident class prediction did not exceed the
rejection threshold and measured the accuracy of the remaining samples.
The accuracy--reject curves were then generated by plotting accuracy as
a function of rejection rate.\looseness=-1

%f5 #&#
\begin{figure}
\centering
\begin{tabular}{@{}cc@{}}

\includegraphics{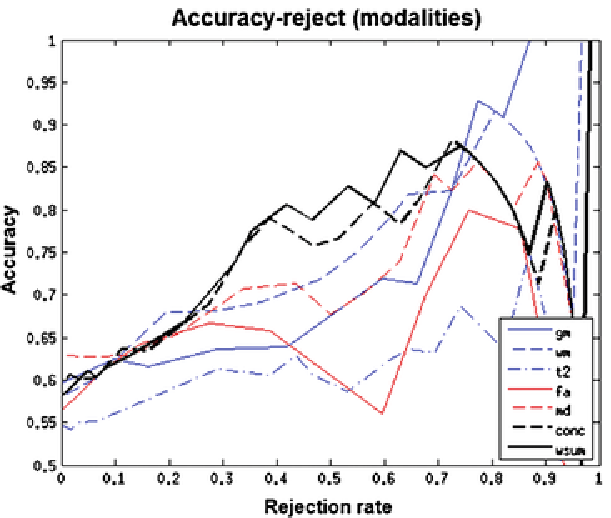}
 & \includegraphics{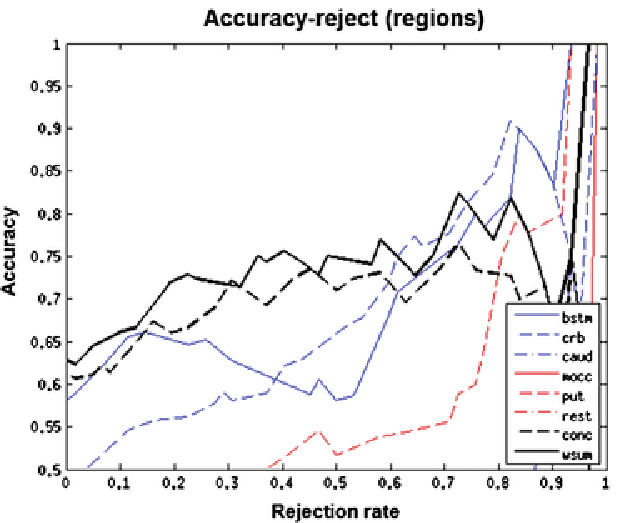}\\
\footnotesize{(a)} & \footnotesize{(b)}
\end{tabular}
\caption{\textup{(a)} Accuracy--reject curve for multi-modality
classifiers. \textup{(b)} Accuracy--reject curve for multi-region classifiers.}
\label{accreject}
\end{figure}

The accuracy--reject curves show that: (i) predictive performance
increases monotonically across most rejection rates and (ii) the
multi-source classifiers perform better than any of the individual
modalities or brain regions across most rejection rates. This implies
that the multi-source classifiers not only make fewer errors, but also
quantify predictive uncertainty more accurately than any of the
individual regions. At high rejection thresholds, the multi-modality
classifier is outperformed by the GM modality, owing to two confident
misclassifications deriving from the FA and MD modalities, suggesting
the possibility of atypical white matter pathology in these subjects.

%s8 #&#
\section{Conclusions}\label{sec8}

In this paper we presented the application of a multinomial logit model
based on GP priors
% fully Bayesian multi-modality multi-class classifier
to the problem of classification of neurological disorders based on
neuroimaging measures.
The proposed model is flexible and highly descriptive, and it can be
employed in scenarios where the focus is on gaining insights into the
relative importance of different data modalities or brain regions in
the application under study.
Also, it allows accurate quantification of the uncertainty in the
predictions, which is crucial in several applications and especially
for predicting disease state in clinical applications.

From a statistical perspective, carrying out the inference task in the
model presented in this paper and in latent Gaussian models in general
represents a serious challenge.
This paper presented a combination of advanced inference techniques
based on MCMC methods that allowed us to tackle this problem in an
efficient way.
Predictions for unseen data were obtained by integrating out all the
parameters in the model, thus capturing the uncertainty in the inferred
parameters.
We also investigated the use of a hyper-prior to integrate out the
choice of the prior. % of An interesting feature of the proposed
%approach is the possibility of inferring sparsity from data, which is
%matter of undergoing investigation and will be presented in other
%reports.

% [afm] The motivating application was based on a set of neuroimaging
%measures of a cohort of 62 patients comprising healthy controls and
%patients affected by a form of Parkinson-plus disorder.
The motivating application for this study aimed to use neuroimaging
measures to classify a cohort of 62 participants, consisting of both
healthy controls and patients affected by one of three variants of
Parkinsonian disorder.
We demonstrated accurate classification of disease state that compares
favorably with the only existing study of which we are aware of
employing whole-brain neuroimaging measures to discriminate between
these disorders [\citet{Focke11}].
{For future work it will be important to: (i) validate how well the
predictive accuracy obtained here generalizes to earlier disease stages
and (ii) investigate methods to improve the predictive accuracy beyond
what was reached here, which will become increasingly important when
the proposed method is evaluated in early stage disease. Construction
of classification features from brain images that better reflect the
underlying pathology may be particularly beneficial in this regard}.
We showed how the results of the inference allowed us to draw
conclusions regarding the relative importance of neuroimaging measures
and brain regions in discriminating between classes.
We also compared the results with simpleMKL, a~nonprobabilistic
multi-modality classifier based on SVMs, which shows lower accuracy
and, most importantly, is not able to address questions regarding the
relative importance of neuroimaging measures and brain regions in a
statistically consistent way.
% [afm] In particular, the proposed method was able to give insights on
%the use of multiple sources for discriminating amongst classes,
%showing that the five modalities considered in this study carry
%similar discriminative information.
% This has important implication as it can be used to guide future
%experiments focusing on one modality only with a great impact on the
%costs associated with the acquisition of new measurements.
In contrast, the proposed method was able to give insights into the
predictive ability of the different neuroimaging sequences, and
suggested that all the modalities investigated in this study carry
similar discriminative information. This has important implications for
planning future studies, and suggests that there is little benefit in
acquiring multiple neuroimaging sequences. Instead, for the purposes of
prediction, acquiring a single structural brain image is probably the
most cost-effective approach.
%Another level of the analysis showed that the proposed method was able
%to detect which brain regions had more importance in the
%discrimination amongst classes.
%Similar to the previous analysis, this study showed that there all
%regions carry some discriminative information, but at the same time
%this analysis seems to indicate that some of them have a higher
%importance in discriminating amongst classes, and this is in line with
%studies in the neuroimaging literature.
Another level of the analysis showed that the proposed method was able
to quantify the predictive ability of different brain regions for
discriminating between classes.
Similar to the previous analysis, this analysis showed that all regions
carry some discriminative information, but at the same time seems to
indicate that some of them have greater predictive ability than others
for different classes. Further, the regional distribution of these
regions is in accordance with the known pathology of the disorders
based on the clinical literature.

\begin{appendix}\label{app}

%s9 #&#
\section{Data acquisition details}

For each subject, a T2-weighted structural image, a T1-weighted spoiled
gradient recalled (SPGR) structural image and a DTI sequence were
acquired using a 1.5T GE Signa LX NVi scanner (General Electric, WI, USA).
All images had whole brain coverage and imaging parameters for the T2
weighted images and DTI sequences have been described previously [\citet
{Blain06}].
Imaging parameters for the SPGR imaging sequence were as follows:
repetition time${} = {}$18~ms, echo time${} = {}$5.1 ms, inversion time${} = {}$450 ms,
matrix size${} = 256 \times152$, field of view (FOV)${} = 240 \times240$.
SPGR Images were reconstructed over a $240 \times240$ FOV, yielding an
in-plane resolution of $0.94 \times0.94$~mm and 124 1.5~mm thick slices.
Subjects provided informed written consent, and the study was approved
by the local Research Ethics Committee.

%s10 #&#
\section{MCMC additional details}
Let $K_{*\cdot}$ be an $m \times(m n)$ block diagonal rectangular
matrix where entries in the $r$th diagonal block contain the covariance
of the test sample with the training samples corresponding to the $r$th
covariance $K_r$.
Also, let $K_{**}$ be an $m \times m$ matrix where the $i,j$ entry is
the covariance of the test sample corresponding to the covariances
$K_i$ and $K_j$.
A priori we assumed zero covariance across latent functions, so
$K_{**}$ will be diagonal.
Using the properties of GPs, given $\fvect$ and $\thetavect$, then
$
p(\fvect_* | \fvect, \thetavect) = \Norm(\fvect_* | \muvect_*,
\Sigma_*)
$
with
$
\muvect_* = K_{*\cdot} K^{-1} \fvect$ and $\Sigma_* = K_{**} -
K_{*\cdot
} K^{-1} K_{\cdot*}
$. Given $N_1$ independent posterior samples for $\fvect$ and
$\thetavect$, we can estimate the integral by
\[
p(\yvect_* | \yvect) \approx\frac{1}{N_1} \sum_{i=1}^{N_1}
\int p(\yvect_* | \fvect_*) p(\fvect_* | \fvect_{(i)},
\thetavect_{(i)}) \,d\fvect_* .
\]
Each of the former integrals can be estimated again by a Monte Carlo
sum, by drawing $N_2$ independent samples from $p(\fvect_* | \fvect_{(i)}, \thetavect_{(i)})$ which is Gaussian:
\[
\int p(\yvect_* | \fvect_*) p(\fvect_* | \fvect_{(i)},
\thetavect_{(i)}) \,d\fvect_* \approx\frac{1}{N_2} \sum
_{j=1}^{N_2} p\bigl(\yvect_* | (\fvect_*)_{(j)}
\bigr) .
\]
The required gradients of the joint log-density follow as
$
\nabla_{\fvect} \mathcal{L} = - K^{-1} \fvect+ \yvect- \pivect
$ and
\[
\frac{\partial\mathcal{L}}{\partial\theta_{cj}} = -\frac{1}{2} \exp[\theta_{cj}] \Tr
\bigl(K_c^{-1} C_j \bigr) + \frac
{1}{2}
\exp[\theta_{cj}] \fvect_c^{\T}
K_c^{-1} C_j K_c^{-1}
\fvect_c + \frac{\partial p(\thetavect_c)}{\partial\theta_{cj}} ,
\]
and the FI for the two groups of variables, along with the negative
Hessian of the priors are
$
G_{\fvect} = K^{-1} + \operatorname{diag}(\pivect) - \Phi\Phi^{\T}
$ and
$
(G_{\thetavect})_{cjr} = \frac{1}{2} \exp[\theta_{cr} + \theta_{cj}] \Tr
(K_c^{-1} C_r K_c^{-1} C_j  ),
$
where $\Phi$ is a $(m n) \times n$ matrix obtained by stacking by row
the matrices $\operatorname{diag}(\pivect_c)$.
The derivatives of the two metric tensors needed to apply RM--HMC can be
computed by using standard properties of derivatives of expressions
involving matrices.
\end{appendix}
% \subsection{Reducing the complexity of applying manifold methods}

% RM--HMC needs the Cholesky decomposition of either $G_{\fvect,
% This means that we would need to factorize the whole $n(m-1) \times
%n(m-1)$ matrix $G_{\fvect, \fvect}$.
% Clearly, this would be feasible only for very special cases;
%fortunately, in our application, this is doable.
% In general, when this is not feasible, so we have a few options as
%described in section 4.
% % The first is to just ignore the off-diagonal blocks, so that the
%Cholesky decomposition of $G_{\fvect, \fvect}$ requires the
%decomposition of $m-1$ matrices of size $n \times n$.
% % The second is to work with block updates of subsets of latent
%functions, whose metric tensor can be computed by selecting the rows
%and columns of $G_{\fvect, \fvect}$ accordingly.

% AOS,AOAS: If there are supplements please fill:
% \sname{Supplement A}
% \stitle{Title}
% \slink[url]{http://lib.stat.cmu.edu/aoas/???/???}
% \sdescription{Some text}

% \bibliographystyle{imsart-nameyear}
% \bibliography{./filippone,./biblio,./additional_refs}

% imsref loaded by akundreckaite, 2012-06-19 09:37:12
%

%suskaldyti doi

\printaddresses

\end{document}